\documentclass[superscriptaddress,twocolumn]{revtex4-2}
\usepackage{graphicx, xcolor, soul}
\usepackage{amsmath,amsthm,amssymb,dsfont}
\usepackage{mdframed}

\usepackage{hyperref}

\begin{document}

\title{Reply to Comment on ``Weak values and the past of a quantum particle''}
\author{Jonte R. Hance}
\email{jonte.hance@bristol.ac.uk}
\affiliation{Department of Quantum Matter, Graduate School of Advanced Science and Engineering, Hiroshima University, Kagamiyama 1-3-1, Higashi Hiroshima 739-8530, Japan}
\affiliation{Quantum Engineering Technology Laboratories, Department of Electrical and Electronic Engineering, University of Bristol, Woodland Road, Bristol, BS8 1US, UK}
\author{John Rarity}
\affiliation{Quantum Engineering Technology Laboratories, Department of Electrical and Electronic Engineering, University of Bristol, Woodland Road, Bristol, BS8 1US, UK}
\author{James Ladyman}
\email{james.ladyman@bristol.ac.uk}
\affiliation{Department of Philosophy, University of Bristol, Cotham House, Bristol, BS6 6JL, UK}

\begin{abstract}
We here reply to a recent comment by Vaidman [\href{https://journals.aps.org/prresearch/abstract/10.1103/PhysRevResearch.5.048001}{Phys. Rev. Res. 5, 048001 (2023)}] on our paper [\href{https://journals.aps.org/prresearch/abstract/10.1103/PhysRevResearch.5.023048}{Phys. Rev. Res. 5, 023048 (2023)}]. In his Comment, Vaidman first admits that he is just defining (assuming) the weak trace gives the presence of a particle---however, in this case, he should use a term other than presence, as this already has a separate, intuitive meaning other than ``where a weak trace is''. Despite this admission, Vaidman then goes on to argue for this definition by appeal to ideas around an objectively-existing idea of presence. We show these appeals rely on their own conclusion---that there is always a matter of fact about the location of a quantum particle. 
\end{abstract}

\maketitle

In his Comment \cite{vaidman2023comment} on our recent paper \cite{Hance2023Weak}, Vaidman seeks to clarify that he does not claim his weak trace approach identified the objectively-existing presence of particles in pre- and postselected scenarios; instead, he claims the weak trace approach defines the ``presence of a quantum particle'' as where it left a weak trace.

 We agree, this would be fine, if the idea of the presence of a particle did not already have a separate, intuitive meaning. This is why we are interested in the idea of the presence of a particle in the first place. 
 
If Vaidman wishes to define some term to mean ``where a particle in a pre- and postselected scenario left a weak trace,'' he is free to do so, but such a term should be free of the implications that terms like ``presence'' possess. The only reason to use a term like ``presence'' is in appeal to some use of this term in another context---such as the conception of presence in classical physics. Therefore, Vaidman either needs to successfully argue that his ``weak trace'' corresponds to our intuitions around notions of ``presence'' (something normally defined either by states being measured as eigenstates of some position/path projection operator, or by appeal to a classical idea of presence), or he should use a different term, or at least clarify that his term refers to something separate to what we intuitively mean by presence.

Despite initially arguing that the weak trace approach does not claim to identify any objectively-existing presence of particles in pre- and postselected scenarios, and just involves defining a weak trace being left along a given path as presence, Vaidman argues one should accept the definition the approach gives for such a presence by directly appealing to ideas around such an objectively-existing idea of presence. For instance, see his statement in the Comment that ``the traces left on the environment that provide evidence of particle interactions have disconnected parts''. This, while used to justify defining particle presence by weak trace, implicitly assumes particles must be present where, and only where, they leave a weak trace---he assumes the very thing he is trying to argue for.

Further, Vaidman's attempt to appeal to our own criteria for using the classical conception of particle presence to rationalise his own approach misses out one key part of our analysis---that there is no need to always assign a particle a localised presence, in a classical fashion, at all times and all locations. Indeed, in some states (e.g., momentum eigenstates) this is by definition impossible according to the laws of quantum mechanics. This is in the same way that, for certain states, there is not a matter of fact about the number of particles in the state (e.g. coherent states).
Our criteria were given as necessary (unless good reason is given) rather than sufficient (especially rather than individually sufficient) to assign particle presence (in a classical fashion).
Vaidman ignores all but one of our criteria, then takes that one remaining criterion as a sufficient condition.
Therefore, Vaidman's argument about our criterion (iii) and our criterion (ii) contradicting, and having to pick one for an approach to identifying the path of a particle, misinterprets our argument.

Vaidman appeals to our criterion (iii)---that (classical) particles interact with other objects and/or fields local to their location. Yet, this does not mean only localised particles interact with other objects and/or fields local to their location, nor that a quantum particle's interaction with another object/field (e.g., the weak trace left on an environment) is sufficient to assign such a classical idea as presence to that particle at that location.

Vaidman comments that ``The fact that the weak value of the velocity of a particle can be larger than the speed of light (see Sec. VIII of \cite{Aharonov1990PropertiesBetween}) does not contradict the special theory of relativity. The experiments involve postselection and their low probability of success prevents a superluminal change in the probability of finding a quantum particle.''

This misunderstands our point, which is that weak values seem to mean something different than standard classical properties, so should not be equated with classical properties. One would expect, by special relativity, anything we consider to be equivalent to the velocity of a particle (such as the propagation speed of a wave) would be limited to being below $c$. Therefore, the fact that these experiments give weak values of velocity greater than $c$, but show nothing which would lead us to question special relativity, that these weak values of velocity must not correspond to true velocities, but instead represent something else.

Vaidman claims ``The weak value approach helps to find quantum protocols which are ``spooky'' if analysed in classical terms.'' However, by ``classical terms” he here means by the definition of particle presence introduced by the weak trace approach. Therefore, the weak trace approach just helps us find quantum protocols which are “spooky” if analysed by the weak trace approach, which seems tautological. Similarly, Vaidman claims the concept of “the local presence of a pre- and post-selected particle defined by the local trace it leaves on the environment” is useful. We are sceptical of this claim, and welcome any evidence that such a definition is in any way useful.

\textit{Acknowledgements -} JRH acknowledges support from Hiroshima University's Phoenix Postdoctoral Fellowship for Research, and the University of York's EPSRC DTP grant EP/R513386/1. JGR and JRH acknowledge support from Quantum Communications Hub funded by EPSRC grants EP/M013472/1 and EP/T001011/1.

\bibliographystyle{unsrturl}
\bibliography{ref.bib}
\end{document}